\documentclass[12pt]{article}
\title{Comment on ``Classical interventions in quantum systems  
II. Relativistic invariance"}
\author{Hrvoje Nikoli\'c \\
Theoretical Physics Division, Rudjer Bo\v{s}kovi\'{c} Institute, \\
P.O.B. 180, HR-10002 Zagreb, Croatia \\
{\normalsize E-mail: hrvoje@faust.irb.hr} \\
\makebox[1in]{} \\
}
\begin{document}
\maketitle
\begin{abstract}
In a recent paper [Phys. Rev. A {\bf 61}, 022117 (2000)],  
A. Peres argued that quantum mechanics is consistent with 
special relativity by proposing that the 
operators that describe time evolution do not 
need to transform covariantly, although the  
measurable quantities need to transform covariantly. 
We discuss the weaknesses of this proposal.  
\end{abstract}
PACS number(s): 03.65.Ta, 03.30.+p 
\vspace{0.5cm}

\noindent
Recently, Peres discussed the role of classical interventions in 
quantum systems \cite{peres1}, with the intention to shed some light 
on one of 
the greatest mysteries of modern theoretical physics: 
the problem of measurement in quantum mechanics. 
In his approach (as well as in the approaches of many others), 
the wave function, described by quantum mechanics, 
is not a material object, but only a 
mathematical tool for calculating probabilities. 
On the other hand, in his approach,    
measurable physical quantities obey classical ontology 
(although not necessarily classical 
deterministic equations of motion). 
In his second paper \cite{peres2}, he argued that unmeasurable
quantities described by quantum mechanics do not need to
transform covariantly under Lorentz transformations, although   
classical, measurable quantities must transform covariantly.  
In this way, he attempted to 
establish a ``peaceful coexistence" of quantum mechanics and 
special relativity. In this comment, we discuss the 
weaknesses of this attempt. 

Let $\rho$ and $\rho_f$ be the density matrices 
that represent the initial and final state, respectively,  
of the same physical system in which measurements are 
performed at the initial and final times. According to 
\cite{peres2}, these two density matrices are related as 
\begin{equation}\label{per2}
\rho_f=\sum_{m,n} K_{mn} \rho K_{mn}^{\dagger} \; .
\end{equation}
The operators $K_{mn}$ are defined in \cite{peres2}. 
For our purpose,  
it suffices to say that they represent a certain generalization 
of the usual unitary operators that describe 
time evolution and that their behavior is described by 
quantum mechanics. The subscripts $m,n$ denote states related 
to the basis vectors that describe the ignored (i.e., not 
measured) part of the physical system \cite{peres1}. 
Eq. (\ref{per2}) refers to a Lorentz frame $S$. In another 
Lorentz frame $S'$, this equation shold be replaced with 
\begin{equation}\label{per4}
\rho'_f=\sum_{m,n} L'_{mn} \rho' {L'}_{mn}^{\dagger} \; .
\end{equation}
According to \cite{peres2}, $\rho$ and $\rho_f$ are physical 
quantities, so they transform covariantly under 
Lorentz transformations. On the other hand, the 
operators $K_{mn}$ and $L'_{mn}$ do not correspond to 
physical quantities, so, according to \cite{peres2}, they 
do not need to transform covariantly. In other words, 
the operator $L'_{mn}$ is not related in any obvious way to the 
operator $K_{mn}$. Below we investigate whether such a proposal 
is consistent. 

Let $\Lambda$ denote the unitary operator corresponding
to the Lorentz transformation. Since $\rho$ and $\rho_f$ 
transform covariantly, we have
\begin{eqnarray}\label{N1}
& \rho' =\Lambda \rho \Lambda^{\dagger} \; , & \nonumber \\ 
& \rho'_f =\Lambda \rho_f \Lambda^{\dagger} \; . & 
\end{eqnarray}
It is easy to show that the compatibility of (\ref{per2}) 
with (\ref{per4}) implies
\begin{equation}\label{N2}
\sum_{m,n} K_{mn} \rho K_{mn}^{\dagger} =
\sum_{m,n} \Lambda^{\dagger} L'_{mn} \Lambda \rho
\Lambda^{\dagger} L'^{\dagger}_{mn} \Lambda \; .
\end{equation}
An obvious way to fulfill Eq. (\ref{N2}) is to propose 
that 
\begin{equation}\label{N3}
K_{mn} = \Lambda^{\dagger} L'_{mn} \Lambda \; .
\end{equation}
This is equivalent to an even more trivial relation
\begin{equation}\label{N4}
K_{mn}=L_{mn} \; ,
\end{equation}
where $L'_{mn}\equiv \Lambda L_{mn}  \Lambda^{\dagger}$.  
If (\ref{N3}) and (\ref{N4}) are fulfilled, then 
quantum mechanics is relativistically covariant. 
However, the basic idea of \cite{peres2} can be expressed as a 
statement that (\ref{N2}) is fulfilled, whereas 
(\ref{N3}) and (\ref{N4}) are not. Therefore, we need to determine 
whether (\ref{N3}) and (\ref{N4}) are 
{\em necessary} consequences of (\ref{N2}). 
To simplify the notation, we write (\ref{N2}) as 
\begin{equation}\label{N2'}
\sum_{A} K_{A}\, \rho\, K_{A}^{\dagger} =
\sum_{A} L_{A}\, \rho\, L^{\dagger}_{A} \; ,
\end{equation}
where $A$ symbolizes the pair $m,n$. 
Note that $K_{A}$ and $L_{A}$ do not depend on the 
choice of the initial state $\rho$. Otherwise, the quantum mechanical 
evolution would not be linear \cite{peres3}.   

Consider first the case in which $A$ can take only one value, 
say $A=1$. Although such a case is not physically realistic, 
no fundamental principle forbids it. In this case, (\ref{N2'}) 
becomes
\begin{equation}\label{N5}
K_{1}\, \rho\, K_{1}^{\dagger} = L_{1}\, \rho\, L^{\dagger}_{1} \; .
\end{equation}
Since $K_{1}$ and $L_{1}$ do not depend on $\rho$ and since 
(\ref{N5}) must be valid for {\em any} $\rho$, it 
follows that $K_{1}=L_{1}$, which corresponds to (\ref{N4}).  
In other words, quantum mechanics must be relativistically
covariant in this case, 
so the claim in \cite{peres2} that a change of the
quantum state related to an EPR setup is instantaneous in
{\em any} frame cannot be true.

When $A$ can take $N>1$ different values, then the 
trivial solution $K_{A}=L_{A}$ of (\ref{N2'}) is not 
the only solution. 
Actually, in this case, there is an infinite number of
solutions and it is not clear (from the equations written 
above) which of them, if not the
trivial one, is the right one. 

In principle, the operators 
$K_{A}$ and $L_{A}$ can be determined uniquely from the 
explicit definitions given in \cite{peres2}. However, these 
definitions involve quantities described by quantum mechanics. 
By assumption of \cite{peres2}, these quantum-mechanical 
quantities are not described by relativistically covariant 
equations. Therefore, it would be a miracle if the 
unique nonrelativistic definitions of $K_{A}$ and $L_{A}$ 
would give the relativistic equation (\ref{N2'}).   
It is not shown in \cite{peres2} that this miracle 
happens. We have shown explicitly that this miracle certainly 
does not happen for $N=1$. 

A possible way out of this problem is to conclude that the definitions 
of $K_{A}$ and $L'_{A}$ given in \cite{peres2} are not really 
unique. In particular, the quantum-evolution laws 
(such as Schr\"{o}dinger equation) are not explicitly written 
in \cite{peres1} and \cite{peres2}. This opens a 
possibility of the existence of an additional 
general principle that determines $K_{A}$ and $L_{A}$ 
in a really unique way. However, although
the case of a large $N$ is realized
in most practical cases, the general principle should be
applicable to all cases, including the case $N=1$.
Therefore,   
the general principle cannot be consistent with the 
assumption that a change of the
quantum state related to an EPR setup is instantaneous in
any frame. Perhaps it is possible to formulate a 
general principle that is consistent with this assumption 
only for large $N$. This suggests that 
noncovariant quantum mechanics 
could be consistent with the covariance of measurable quantities    
only in the large-$N$ limit, which gives a statistical, approximate 
status to the proposal in \cite{peres2}. However, 
there is no much use of this proposal without an 
explicit general principle that determines the unique  
nontrivial solution of (\ref{N2'}).     

Note that in \cite{peres2} 
Peres derived certain consistency conditions that provide 
that (\ref{per2}) is consistent when $\rho$ and 
$\rho_f$ refer to measurements performed at 
(almost) the same time. However, these consistency 
conditions do not imply that (\ref{N2}) can be true  
without (\ref{N3}) being true. Indeed, these consistency 
conditions are not in contradiction with (\ref{N3}).  

To summarize, the definition of $K_{mn}$ and $L_{mn}$ in 
\cite{peres2} is either unique or ununique. If it is unique, 
then it is not clear how it can be consistent with 
(\ref{N2}). If it is ununique, then it can be chosen such 
that it is consistent with (\ref{N2}) without (\ref{N3}) being true, 
but then the ununiqueness is a problem by itself. In both cases, 
if $N=1$, then (\ref{N2}) cannot be true without (\ref{N3}) being true.

It is not the intention of this comment to solve the 
problem of measurement in 
quantum mechanics and its consistency with special 
relativity. However, we note that, in our opinion, 
there are two promising types of approaches to 
the resolution of this problem. One is to generalize 
quantum mechanics by a nonlinear theory, as, 
for example, in \cite{wein}. In particular, in this case,  
$K_{mn}$ may depend on $\rho$, so, 
even for $N=1$, (\ref{N2}) may be 
consistent without (\ref{N3}) being true. 
In the second type of approaches, the  
linear Schr\"{o}dinger equation is exact, but there 
exists a preferred coordinate frame, which violates  
the principle of relativity. For example, the  
de Broglie--Bohm interpretation of quantum field 
theory requires a preferred coordinate frame \cite{hol}.   

I am grateful to A. Peres for the discussion on his work, 
although we do not share the same point of view on this subject. 
This work was supported by the Ministry of Science and Technology of the
Republic of Croatia under Contract No. 00980102.

\end{document}